\documentclass{iopjournal}
\usepackage{amsmath}
\usepackage[numbers,sort&compress]{natbib}
\usepackage{ragged2e}

\begin{document}
\justifying
\articletype{Paper} 

\title{Localization from Infinitesimal Kinetic Grading: Finite-size Scaling, Kibble–Zurek Dynamics and Applications in Sensing}

\author{Argha Debnath\orcid{0000-0001-7131-6043}, Ayan Sahoo\orcid{0009-0006-4454-9742} and Debraj Rakshit$^{*}$\orcid{0000-0002-5227-0972}}

\affil{Harish-Chandra Research Institute, A CI of Homi Bhabha National Institute,  Chhatnag Road, Jhunsi, Prayagraj 211 019, India}

\affil{$^*$Author to whom any correspondence should be addressed.}

\email{debrajrakshit@hri.res.in}


\begin{abstract}
\justifying

We study a one-dimensional lattice model with site-dependent nearest-neighbor hopping amplitudes that follow a power-law profile. The hopping variation is controlled by a grading exponent, $\alpha$, which serves as the tuning parameter of the system. In the thermodynamic limit, the ground state becomes localized in the limit $|\alpha| \to 0$, signaling the presence of a critical point characterized by a diverging localization length. Using exact diagonalization methods, we perform finite-size scaling analysis, and extract the associated critical exponent governing the near-critical behavior. To further characterize the criticality, we analyze inverse participation ratio (IPR),  energy gap between the ground and first excited state, and fidelity-susceptibility.  We also investigate the non-equilibrium dynamics by linearly ramping the hopping profile at various rates and tracking the evolution of the localization length and the IPR. The Kibble-Zurek mechanism successfully explains the resulting dynamics of the system via the critical exponents obtained from static scaling analysis. Beyond its fundamental significance, the kinetic-grading-induced localization transition provides a natural platform for quantum sensing. Using the critical enhancement of the quantum Fisher information (QFI), we demonstrate that the system enables quantum-enhanced parameter estimation of the grading exponent. We propose both adiabatic and dynamical quantum critical sensors and demonstrate that they exhibit enhanced scaling of the QFI. Our results therefore establish graded kinetic systems not only as a new setting for localization physics, but also as a potential resource for designing quantum-enhanced sensing devices.

\end{abstract}

\section{\label{sec:intro}\textcolor{red}{Introduction}}

Many recent studies have explored the idea of quantum sensors. Their operation is based on the properties of many-body quantum systems at criticality. Several of these works
have recognized quantum phase transitions, such as first and second-order quantum phase transitions, and localization
transitions as useful quantum resources that can be exploited
for engineering the critical quantum sensors \cite{zanardi2008quantum,degen2017quantum, rams2018limits,di2023critical, mondal2024multicritical, sahoo2024enhanced,bayat20251,agarwal2025critical,ghosh2026journey}. It is difficult to measure the strength of an unknown weak field. The challenge becomes greater when high precision is required. A recent work has
proposed a weak field sensing device by employing Stark
weak field inducing localization \cite{he2023stark}.
At the quantum critical (or transition) point, the fidelity susceptibility follows a power-law scaling with the system size, $L$ as $L^{2/d\nu}$,
where $d$ denotes the spatial dimension of the system, while $\nu$ is the critical exponent that characterizes the divergence of the localization (or correlation) length near the transition \cite{sahoo2024enhanced}. The localization induced by disorder, proposed by Anderson, is a longstanding research topic in condensed-matter physics \cite{anderson1958absence}. Sufficiently strong disorder suppresses quantum transport, leading to the confinement of the wavefunction within a localized region of the lattice. Since then, numerous models have explored different ways of introducing randomness to drive the transition between extended and localized phases. Beyond truly random potentials, the Aubry–Andr{\'e} model demonstrated that a quasi-periodic modulation can also induce localization through self-duality \cite{aubry1980analyticity,albert2010localization,PhysRevA.109.L030601}. A third prominent route is Stark localization: a uniform field applied across the lattice suppresses tunneling and leads to spatially localized single-particle states even in the absence of disorder \cite{wannier1960wave, fukuyama1973tightly,kolovsky2003bloch,kolovsky2008interplay,kolovsky2013wannier, pwfy-2j1v, vqnf-zd5w, PhysRevB.111.L220201, PhysRevB.111.174202, PhysRevB.111.174209}. 
Apart from that, many-body localization in the presence of interaction has attracted a lot of interest in research \cite{wu2019bath,schulz2019stark,van2019bloch,bhakuni2020drive,taylor2020experimental,wei2022static}. These theoretical advances have inspired diverse experimental realizations in ultra-cold atoms, photonic lattices, and solid-state platforms \cite{billy2008direct,roati2008anderson,deissler2010delocalization,schreiber2015observation,morong2021observation,guo2025observation}. Quantum effects can be utilized to improve measurement sensitivity, making it possible to develop a new generation of quantum sensors.

This motivates a systematic investigation of how an infinitesimal gradation in hopping amplitudes can induce localization, and how the resulting localization transition can be characterized in terms of critical exponents and dynamical universality classes—issues that remain largely unexplored. Non-uniform or graded hopping models have been studied in a variety of contexts, including optical lattices and photonic waveguide arrays, where spatially inhomogeneous couplings are known to generate localized modes \cite{lima2004finite,cheraghchi2005localization,de2010absence,dias2010kosterlitz,assunccao2011coherent,deng2012delocalization,saha2023localization,tabanelli2024reentrant}. These developments naturally raise several fundamental questions: How does localization emerge under a generic power-law hopping profile? Does a delocalization–localization crossover arise in finite-size systems as the exponent $\alpha$—which controls the strength of the hopping gradation—is varied? More intriguingly, can a small drift from uniform hopping, i.e., $|\alpha| \to 0$, suffice to localize the ground state in the thermodynamic limit? Beyond these foundational aspects, it is also pertinent to ask whether such localized states, induced purely by hopping inhomogeneity, can serve as useful quantum resources—for instance, in enhancing sensitivity for quantum sensing applications.

For $\alpha=1$, the hopping amplitudes grow linearly along the chain, leading to strongly localized eigenstates reminiscent of Stark localization but generated via a graded kinetic term rather than a diagonal potential. Here we treat this case not as an isolated, previously-known limit, but as a particular representative of a broader class of graded hopping models parametrized by $\alpha$, and we focus on how localization emerges as $|\alpha| \to 0$. A key structural feature of the model is that for any nonzero grading exponent $\alpha>0$, the hopping strengths grow boundlessly with the
site index, and the local kinetic energy scale therefore becomes non-extensive in the thermodynamic limit, while it vanishes for $\alpha<0$. This makes the limit $\alpha\to 0$ intrinsically singular: while $\alpha=0$ corresponds to a uniform, translationally invariant and has eigenstates as fully delocalized states, an arbitrarily small $\alpha$ produces qualitatively different kinds of kinetic landscapes capable of supporting localization. This sharp distinction motivates a careful study of how localization onsets as $\alpha$ departs from zero.  An intriguing question in this context is if the same scaling exponents may describe the criticality when approached from the either side determined by the sign of $\alpha$.

We undertake a detailed characterization of how localization
emerges under a power-law grading of the hopping amplitudes. By examining the ground-state wavefunction, inverse participation ratio, and fidelity susceptibility, we find a smooth delocalization–localization crossover in finite systems, which sharpens with increasing system size and develops into a sharp critical point in the thermodynamic limit. A distinctive feature of the
model is that the delocalized regime collapses to the single point $\alpha=0$: the
uniform lattice is extended, whereas any infinitesimal grading $|\alpha|>0$
eventually produces localized ground states. The transition therefore consists of a
sharp critical point at the boundary between a measure-zero delocalized point and a
one-sided localized phase. We can introduce a critical point, namely $\alpha_c$ where the localization length, $\xi$, diverges as $\xi\sim|\alpha-\alpha_c| ^{-\nu}$. Also, a vanishing energy gap, and consistent finite-size scaling of
multiple observables, enable us to extract the associated critical exponents.

 The static analysis presented above is complemented by a dynamical study. A central objective of this work is to examine non-equilibrium processes initiated from the ground state deep within the localized phase and subsequently driven across the transition point. The system experiences a critical slowing down, which leads to an impulse regime near the critical point and the system can no longer adiabatically follow the instantaneous ground state. The quantum Kibble–Zurek mechanism (KZM) \cite{dziarmaga2010dynamics,polkovnikov2011colloquium,kibble2007phase,kibble1976topology,zurek1985cosmological,zurek1996cosmological} in its quantum extension describes how systems deviate from adiabatic evolution, which provides a universal framework for describing defect formation, scaling laws, and non-adiabatic dynamics in the vicinity of continuous phase transitions. The KZM framework has been proven to be effective in quantum phase transition studies through its application to systems that cross a single quantum critical point under linear or nonlinear ramps \cite{zurek2005dynamics,dziarmaga2005dynamics,polkovnikov2005universal,damski2007dynamics,schutzhold2006sweeping,uhlmann2007vortex,sen2008defect,bermudez2009topology,deng2009dynamical,uhlmann2010n,uhlmann2010system,chandran2013kibble}. Within this approach, the non-equilibrium dynamics near criticality follow universal Kibble–Zurek scaling (KZS), which provides an independent dynamical verification of the critical exponents extracted from the static analysis. Recently, increasing attention has turned toward extending this paradigm to the localization transitions \cite{huang2014kibble,roosz2014nonequilibrium,morales2014resonant,serbyn2014quantum,liu2015universal,sinha2019kibble,bu2022quantum,bu2023kibble,liang2024quantum,wang2025driven}. Such studies are gaining importance not only for the advancement of theoretical understanding, but also due to their experimental relevance and  potential applications in emerging quantum technologies \cite{meldgin2016probing,anquez2016quantum,clark2016universal,chiu2018quantum}. We carried out an in-depth investigation which reveals phase transitions among localized phase and extended phase through KZM. We show dynamical scaling exponents from the KZS analysis matches with the critical exponents obtained from the static scaling analysis.

Based on the study of quantum criticality,  we show that the
system under study has important applications in quantum
metrology \cite{giovannetti2006quantum}. We demonstrate that the proposed probe provides enhanced quantum sensing over a broad region of the extended phase up to the critical point in the single-particle regime, making it suitable for detecting weak fields with high precision. We also investigate a dynamical sensing approach in which a sudden quench is used to encode the unknown parameter $\alpha$ into the probe state. Compared with adiabatic methods, dynamical sensing offers several practical benefits. The evolution time can be used as an additional resource to improve the estimation precision, and the protocol avoids the limitations associated with critical slowing down. In addition, sudden-quench schemes are often easier to implement experimentally than the finite-time adiabatic ramps required in conventional sensing protocols.

The organization of the paper is as follows: In Sec.~\ref{sec:model}, we
describe our Hamiltonian comprising only 
with modulated hopping. In Sec.~\ref{sec:qc}, we set the background of localization length, inverse participation ratio, fidelity susceptibility. We study the localization properties with modulated hopping co-efficient. We study critical properties of the localization transition by extracting critical exponents for finite system sizes. In Sec.~\ref{sec:kzs}, we investigate quantum criticality through dynamical driving to study the KZS. We went on investigating quantum sensing applicability in Sec.~\ref{sec:fisher}. In Sec.~\ref{sec:conc}, we summarize our results.

\textcolor{red}{\section{\label{sec:model}Model} }
 We consider a one-dimensional tight-binding lattice with site-dependent nearest-neighbor hopping amplitudes following a power-law profile. The Hamiltonian of the system is
\begin{eqnarray}\label{eq:main}
    \hat{H} = -\sum_{i=1}^{L-1} t_i\left(\hat{c}_i^{\dagger} \hat{c}_{i+1} + \hat{c}_{i+1}^{\dagger} \hat{c}_i\right),
\end{eqnarray}
Here, the hopping amplitude is $t_i=J i^{\alpha}$, with  $i$ being a positive bond index connecting sites $i$ and $i+1$.  For $\alpha=0$, all hopping amplitudes become equal, and the model reduces to the uniform nearest-neighbor tight-binding chain. The exponent 
$\alpha$ serves as a control parameter that governs the degree of gradation, thereby introducing a spatially varying nonlinearity in the hopping term. Here, $\hat{c}_i^{\dagger}(\hat{c}_i)$ is the fermionic creation (annihilation) operator at the $i^{th}$ site. We impose open boundary condition and investigate finite-size systems by employing the exact diagonalization technique considering $J=1$.

We perform exact diagonalization for solving Eq.~\ref{eq:main}. The Hilbert space is spanned by the occupation-number basis and for  a general L-site lattice with one particle the basis is generated via $|i\rangle = \hat{c}_i^{\dagger}|0\rangle$, where $|0\rangle$ is the vacuum state (lattice with no particle), and $i=1,2,3, \cdots$.  Here {$|i\rangle$} denotes the single-particle computational basis--the standard site basis of the one-dimensional lattice. Apart from direct diagonalization studies, useful insight can also be obtained from continuum envelop form of Eq.~\ref{eq:main}, which we briefly mention in the Appendix.

The point $\alpha = 0$ corresponds to the uniform lattice with translational symmetry, where all single-particle states are completely delocalized. However, any nonzero $\alpha$, no matter how small, breaks translational invariance and effectively introduces a spatial bias in the hopping landscape. The Hamiltonian contains no on-site terms and consists solely of nearest-neighbor hopping between alternating lattice sites. As a result, it admits a sublattice operator
 $\hat{S} = \sum_i (-1)^i |i\rangle\langle i|$ that anti-commutes with the Hamiltonian, $\{\hat{S} , \hat{H} \} = 0$. This immediately implies a spectral mirror symmetry: every eigenstate $|\psi\rangle$ of energy $E$ has a partner $ \hat{S} |\psi\rangle$ with energy
$-E$. Consequently, all nonzero eigenvalues appear in $\pm E$ pairs, and for lattices of odd size a single zero-energy mode must exist.
This structural property is intrinsic to off-diagonal tight-binding models and holds for all values of the grading exponent $\alpha$.

The graded tight-binding Hamiltonian introduced above provides a clean,
disorder-free setting in which localization emerges solely from a controlled variation
of the hopping amplitudes. In
the following sections, we analyze this behavior by examining the ground state
localization properties, finite-size scaling, and associated critical exponents, and
we subsequently explore the real-time critical dynamics within the Kibble--Zurek
framework.

\textcolor{red}{\section{\label{sec:qc}Localization Transition And Finite-Size Scaling}}

 The system is fully delocalized at $ \alpha=0$. To understand how the graded hopping profile affects the delocalized state, it is essential to examine its dependence on system size. In order to distinguish the different quantum phases and characterize the criticality, we analyze several key diagnostic quantities, such as localization length, inverse participation ratio, energy gap and fidelity susceptibility. In the following, we study the ground-state behavior of these quantities as the grading exponent $\alpha$ is varied for different system sizes.

\subsection{\label{subsec:localength}\textnormal{\textbf{Localization Length}}}

A second-order quantum phase transition is characterized by a diverging length scale of the correlation function. Although the localization transition considered here does not belong to the Landau paradigm, an analogous scaling framework can be formulated in which the localization length plays a role similar to that of the correlation length. 

In the present model, the spatial variation of the hopping amplitudes naturally leads to ground states whose support may concentrate near a site. This confinement is naturally quantified by the localization length, which measures how quickly the wave function decays away from its peak. A finite localization length implies that the state is truly localized and largely insensitive to the system size, whereas a diverging localization length signals the approach to a delocalized or critical regime. Thus, tracking how the localization length varies with the grading exponent $\alpha$ and with system size offers a direct means of identifying the localization-delocalization crossover in the finite-size systems.

To define localization length, $\xi$, we take into account root-mean-square width of the ground state  \cite{liang2024quantum,sahoo2025stark}
\begin{equation}
\xi=\sqrt{\sum_i^L(i-i_c)^2p_i},    
\end{equation}
where $i$ denotes the lattice site, $p_i(=|\psi(i)|^2)$ is the single-particle probability density at site $i$ with $\psi(i)$ being the ground state, and $i_c$ is the localization center having
the expression $i_c=\sum_i^Lip_i$. At the thermodynamic limit, $\nu$ governs the degree of divergence as control parameter $g$ nears critical point $g_c$ as
\begin{equation}
\xi\propto |g-g_c|^{-\nu},
\end{equation}
 
\begin{figure}
\centering
\includegraphics[scale=0.4]{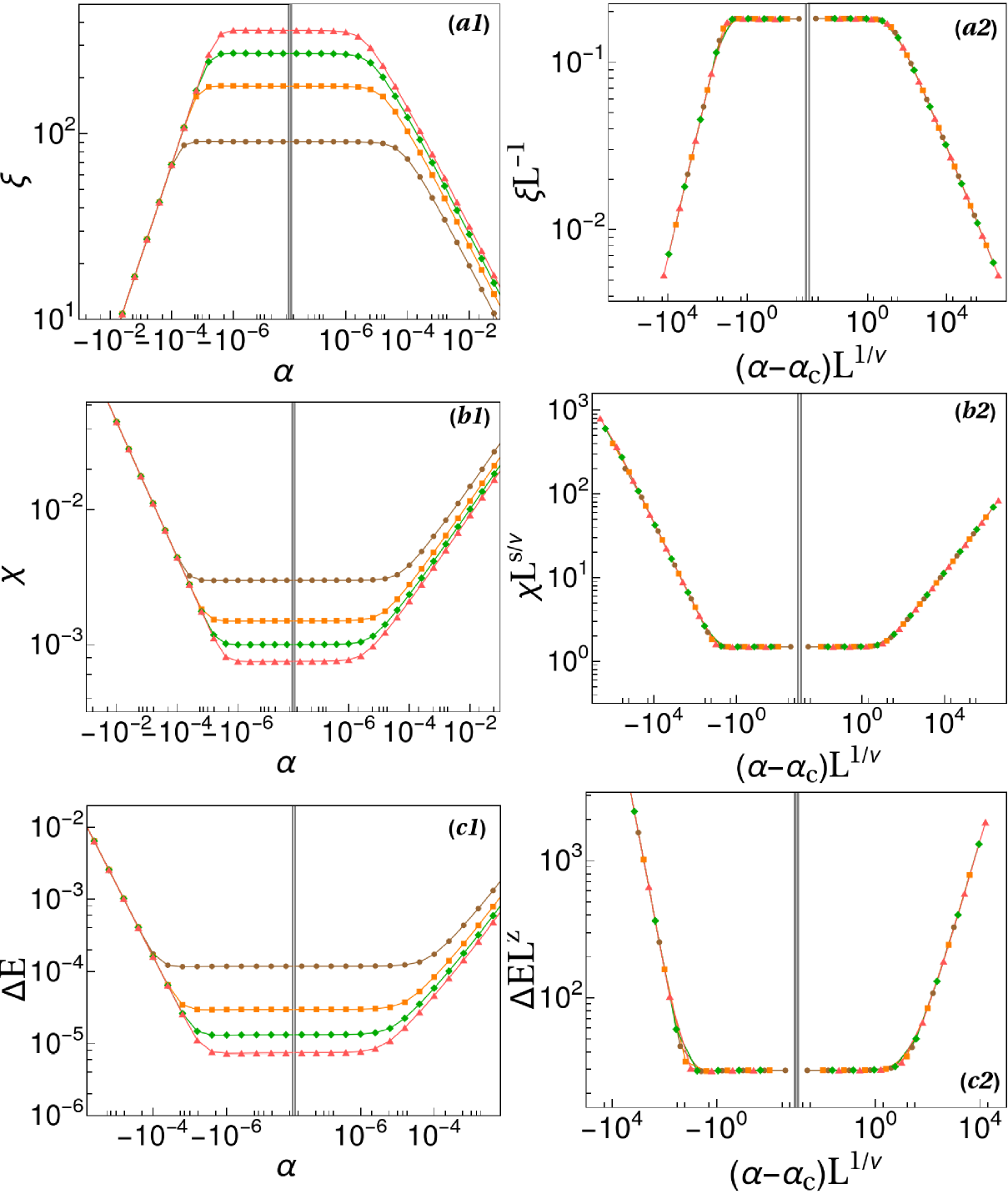}
\caption{Variation of (a1) localization length($\xi$), (b1) IPR($\chi$), and (c1) energy gap($\Delta E$) with  $\alpha$ for different system sizes, $L=500$ (yellow circle), 1000 (orange square), 1500 (green diamond), 2000 (pink up-triangle) for ground state using exact diagonalization. Collapse plot with $(\nu,s,z)\sim(0.49(1),0.49(1),2.02(2))$ are shown for (a2) $\xi$, (b2) $\chi$, (c2) $\Delta E$ with their re-scaled axis. $\alpha_c\sim 10^{-9}$.}\label{fig:loc}
\end{figure}

We study the ground state and analyze the dependence of $\xi$ on  $\alpha$ for several system sizes $L$ in Fig.~\ref{fig:loc}(a1). This figure illustrates that for finite-sized systems, $\xi$ remains constant for very small $\alpha$ maintaining it's delocalized state. But the wavefunction progressively localizes after a certain threshold of $\alpha$, say $\alpha_T$ for every finite $L$. The above analysis suggests that the apparent threshold $\alpha_T(L)$ extracted from finite-size data drifts rapidly toward zero as $L$ increases. Extrapolating this trend indicates that the limiting value $\alpha_c \equiv \lim_{L\to\infty}\alpha_T(L)$ lies extremely close to zero, with numerical estimates yielding a small finite value of order $10^{-9}$,
which is essentially set by the numerical resolution and the finite system sizes considered in this work. 

We perform data collapse for determining the scaling exponent $\nu$ via the cost function approach (see subsection \ref{subsec:cost}). It offers an unbiased and robust method for extracting the scaling exponents, independent of fitting ambiguities. At the same time, it is an effective approach for overcoming finite-size artifacts. In order to obtain data collapse, we use the following scaling ansatz:
\begin{equation}
\xi/L=f_1\left((\alpha-\alpha_c) L^{1/\nu}\right),
\label{eq:loca_an} 
\end{equation}
where $f_1[.]$ is an arbitrary function. Fig.~(a2) displays the data collapse, where we have rescaled the quantities $\xi$ by  $\xi/L$ and $\alpha$ as $(\alpha-\alpha_c)L^{1/\nu}$. The best data collapse is obtained for $\nu=0.49(1)$. Noticeably, despite qualitatively different kinds of kinetic landscapes on either side of $\alpha=0$, the divergence of the localization length at criticality is governed by the same scaling exponent, which is essentially insensitive to the sign of $\alpha$.

It is instructive to contrast the ground state critical behavior of our model with the well-established one-dimensional localization universality classes. In the 1D Anderson model with weak diagonal disorder, the localization length at the unperturbed band edge scales with the disorder strength as $\xi \propto W^{-2/3}$ in the limit $W \to 0$  \cite{cestari2011critical}. In the Aubry–Andr{\'e}–Harper model the exponent associated with ground state is $\nu = 1$ \cite{PhysRevA.109.L030601}. In the pure Stark (Wannier–Stark) problem, the band-edge scaling is governed by the Airy equation, giving rise to a much smaller exponent $\nu = 1/3$. By contrast, the graded-hopping model studied here exhibits a ground-state localization-length exponent $\nu \simeq 1/2$, placing it in a universality class distinct from Anderson, Aubry–Andr{\'e}, and Stark localization.

\subsection{\label{subsec:ipr}\textnormal{\textbf{Inverse Participation Ratio}}}  IPR, $\chi$, is the next observable that we use for probing the delocalization-localization transition. While the localization length characterizes how rapidly a wavefunction decays away from its localization center, the IPR quantifies the distribution of the weight of the wavefunction across the lattice. It is defined as
\begin{equation}
\chi=\sum_i^Lp_i^2. 
\end{equation}
We plot the variation of the ground state IPR value with the increase of $\alpha$ for different $L$ in Fig.~\ref{fig:loc}(b1). For a particular $L$, $\chi$ remains constant for small $\alpha$ below a threshold. In this delocalize region $\chi$ in inversely proportional to $L$.  Beyond this region, the $\chi$ loses its $L$ dependence and enters in the localize regime. Close to the critical point, IPR follows a power-law dependence on the system size, $\chi \propto L^{-s/\nu}$. In the thermodynamic limit, it also varies with the distance from the critical point as $\chi \propto |g-g_c|^{s}$, where $s$ is the corresponding critical exponent \cite{sahoo2025stark}. On the other hand, near criticality, the IPR scales . We consider the following ansatz for extracting the corresponding scaling exponent: 
\begin{equation}
\chi=L^{-s/\nu}f_2\left((\alpha-\alpha_c) L^{1/\nu}\right), \label{eq:ipr_an} 
\end{equation}
where $f_2[.]$ is an arbitrary function. By rescaling $\chi$ and $ \alpha$ as $\chi L^{s/\nu}$ and $(\alpha-\alpha_c) L^{1/\nu}$, respectively, according to Eq.~\ref{eq:ipr_an}, we find that for $s=0.49(1)$, for which we obtain the best data collapse shown in Fig.~\ref{fig:loc}(b2) using the cost function.

\subsection{\label{subsec:energygap}\textnormal{\textbf{Energy Gap}}} 
The energy gap, $\Delta E$,  between the first excited state and the ground state can also be used to characterize the criticality. Fig.~\ref{fig:loc}(c1) shows $\Delta E$ with hopping strength $\alpha$ for different $L$. $\Delta E$ goes to zero as we increase $L$ for small $\alpha$ proving $\Delta E \propto L^{-z}$. Energy gap follows same trend of delocalization-localization transition with increasing $\alpha$ as $\xi$ and $\chi$. We extract exponent related to $\Delta E$ using the following ansatz,
\begin{equation}
\Delta E=L^{-z}f_3\left((\alpha-\alpha_c) L^{1/\nu}\right),\label{eq:eng_an} 
\end{equation}
where $f_3[.]$ is an arbitrary function.
Using the above ansatz, we obtain data collapse and plotted in Fig.~\ref{fig:loc}(c2), with their corresponding scaled axis. Finally, the obtained exponents are $\{\nu, s, z\}=\{0.49(1),0.49(1),2.02(2)\}$.

\subsection{\label{subsec:cost}\textnormal{\textbf{Cost Function}}}
\begin{figure}
\centering
\includegraphics[scale=0.5]{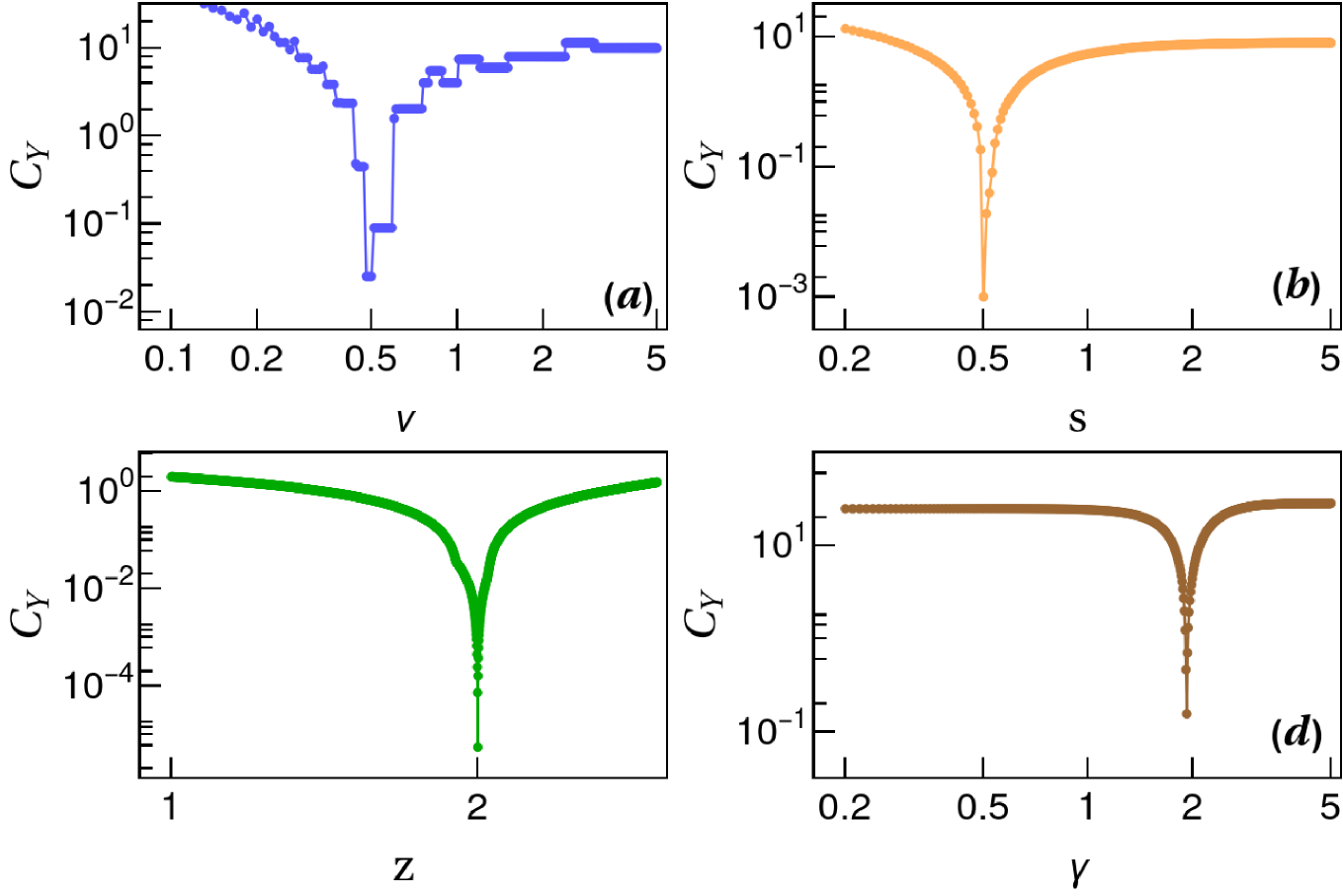}
\caption{Cost function to find $\nu$, $s$, $z$ and $\gamma$. A global minimum of
$C_Y$ is at (a) $\nu=0.49(1)$, (b) $s=0.49(1)$, (c) $z=2.02(2)$, (d) $\gamma=1.96(4)$.}\label{fig:cost}\end{figure}

 In order to find the best fit values of the scaling exponents we take into account cost function, $C_Y$ is defined as \cite{vsuntajs2020ergodicity,modak2021finite}
\begin{equation}
    C_Y=\frac{\sum_{i=1}^{N-1}|Y_{i+1}-Y_i|}{max\{Y_i\}-min\{Y_i\}}-1,   
    \label{eq:cost}
\end{equation}
We create total $N$ data points, $\{Y_i\}$, for different values of $\alpha$ from each $L$. All $N$ values of $|Y_i|$  are sorted with the rising value
of $Lsgn[\alpha-\alpha_c] [\alpha-\alpha_c]^\nu$ \cite{sahoo2025stark}. To obtain perfect data collapse $C_Y=0$ is needed. Instead of zero, a global minimum value of $C_Y$ can also provide approximate scaling exponent. Exponents are extracted as $C_Y$  at $\nu=0.49(1)$, $s=0.49(1)$, $z=2.02(2)$ and $\gamma=1.96(4)$ show a global minima (see Figs.~\ref{fig:cost}); data collapse is achieved with these values.

\textcolor{red}{\section{\label{sec:kzs}Kibble-Zurek Scaling Of Driven Dynamics}}
 We now turn to investigate Kibble–Zurek scaling (KZS) exploiting the phase transition in our model \cite{liang2024quantum,sun2025}. The KZM establishes a unified framework that explains how systems produce excitations when they experience continuous phase transitions at a controlled speed. The system maintains adiabaticity when the control parameter remains distant from the criticality. The system generates unavoidable excitations because the essential energy gap approaches zero as it approaches the critical point. The KZM describes how adiabaticity fails during a process which starts from an easily accessible ground state before reaching a complex final state. We take our initial state as a localized one and  linearly vary $\alpha$ across localization-delocalization transition point with speed $R$. The time evolution of $\alpha$ is given by \cite{liang2024quantum,sun2025}
\begin{eqnarray}
\alpha(t)=\alpha_0+Rt,\label{eq:kz}
\end{eqnarray}
where $\alpha_0$ represents the control parameter of the initial state at $t = 0$. The KZS framework dictates that adiabaticity can be maintained when the condition  $|\alpha|>R^{1/r\nu}$ is satisfied, with the scaling exponent $r=z+1/\nu$ \cite{liang2024quantum}. In this regime, the Hamiltonian evolves slowly enough that the system remains aligned with the instantaneous ground state throughout the evolution. By contrast, when $|\alpha|<R^{1/r\nu}$, the intrinsic response of the system becomes slower than the external driving rate, signaling its entry into the impulse regime, where the evolution effectively freezes and excitations are generated.

\begin{figure}
\centering
\includegraphics[scale=0.4]{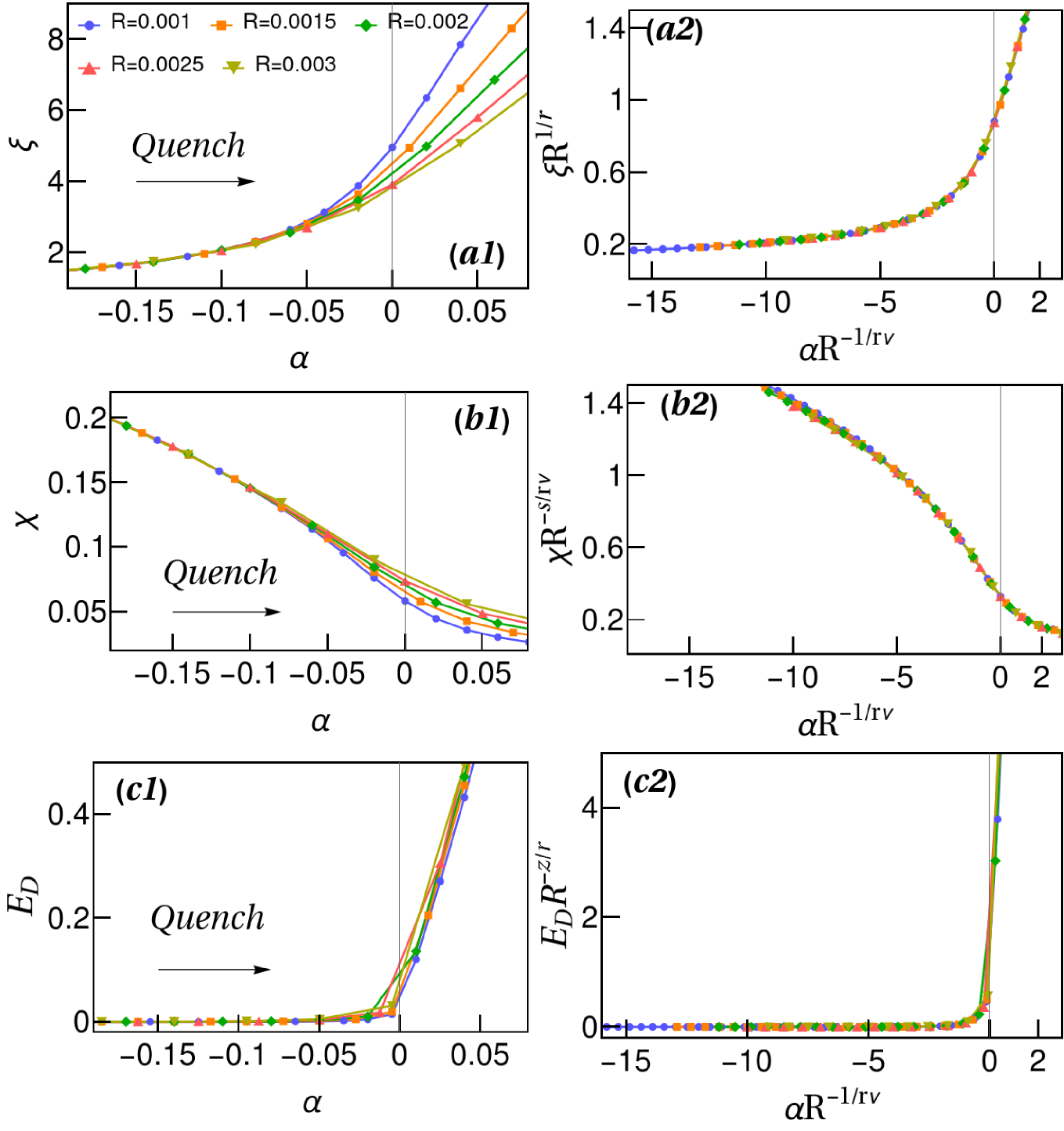}
\caption{Driven dynamics of the ground state. The curves of $\xi$ versus $\alpha$ for $R=0.001$ (blue circle), $R=0.0015$ (orange diamond), $R=0.002$ (green square), $R=0.0025$ (pink up-triangle), $R=0.003$ (yellow down-triangle) (a1) before
and (a2) after rescaled with $R$. The curves of $\chi$ versus $\alpha$ (b1)
before and (b2) after rescaled with $R$. We have plotted $E_D$ calculated from Eq.~\ref{eq:kz_an3} versus $\alpha$ (c1) before and (c2) after rescaled with $R$. The arrows in (a1), (b1)
and (c1) point the quench direction. The lattice size is $L=1000$.}\label{fig:kz}
\end{figure}

To see how $\xi$ and $\chi$ behave around criticality while driven dynamically, we introduce following ansatz,
\begin{equation}
\xi=R^{-1/r}f_4\left(\alpha R^{-1/r\nu}\right), \label{eq:kz_an1}
\end{equation}
\begin{equation}
\chi=R^{-s/r\nu}f_5\left(\alpha R^{1/r\nu}\right), \label{eq:kz_an2}
\end{equation}
where $f_4[.]$ and $f_5[.]$ are arbitrary functions. Throughout this Kibble-Zurek dynamical work, we use a system size of $L=1000$, which is sufficiently large for finite-size effects to be safely neglected in the real-time simulations. We take ground state at $\alpha_0=-0.5$ as our initial state. Fig.~\ref{fig:kz}(a1) shows the evolution of
 $\xi$ for several $R$. At initial stages of evolution, our initial localized state with a small localization length $\xi$ is almost insensitive to the ramp rate $R$, allowing the dynamics to closely follow the instantaneous ground state in an effectively adiabatic manner. However, as $\alpha$ approaches to the transition point, the trajectories differ from one another depending on their ramp rates $R$, signaling the onset of the impulse regime where the evolution can no longer remain adiabatic. We rescale $\xi$ and $\alpha$ as $\xi R^{1/r}$ and $\alpha R^{-1/\nu r}$, respectively, according to the ansatz in Eq.~\ref{eq:kz_an1}.  We collapse the rescaled curves onto each other using the same scaling exponents from the static case as shown in Fig.~\ref{fig:kz}(a2). Similarly, Fig.~\ref{fig:kz}(b1) shows the evolution of $\chi$ for different
$R$. In contrast to the localization length, the fidelity susceptibility starts from a large initial value and evolves through an adiabatic stage in which its behavior is nearly independent of the ramp rate 
$R$. As the system nears the critical region, however, it exhibits a pronounced impulse response, reflecting the breakdown of adiabaticity. After
rescaling $\chi$ and $\alpha$ as $\chi R^{-s/\nu r}$ and $\alpha R^{-1/\nu r}$ according to Eq.~\ref{eq:kz_an2}, respectively, we get collapse of the rescaled curves onto each other using the same scaling exponents from the static case as shown in Fig.~\ref{fig:kz}(b2).

In KZM, near the transition point the state fails to follow the instantaneous change of the Hamiltonian and the deviation in terms of energy can be written as \cite{zhai2022nonequilibrium}
\begin{equation}
E_D\equiv Re\left[\langle\psi(t)|H(t)|\psi(t)\rangle-E_g(t)\right], \label{eq:kz_an3}
\end{equation}
where $H(t)$ is the instantaneous Hamiltonian in the driving process, and $E_g (t)$ is the corresponding ground state energy. $\psi(t)=e^{-iH(t)t}\psi(0)$, and $\psi(0)$ is the initial state. Away from critical point $E_D(t=0)=0$, and remains so; however, near criticality the energy obtained by driving $\psi(0)$ with $H(t)$ does not resemble the ground state energy of the instantaneous Hamiltonian obtained through $E_g(t)$ shown in Fig.~\ref{fig:kz}(c1). This can also be attributed to the excitations generated in the ground state as it nears the transition point. The driven dynamics satisfy the following ansatz:
\begin{equation}
E_D=R^{-z/r}f_6\left(\alpha R^{-1/r\nu}\right), \label{eq:kz_an4}
\end{equation}
where $f_6[.]$ is an arbitrary function. Then, by rescaling $E_D$and $\alpha$ as $R^{z/r}$ and
$\alpha R^{-1/r\nu}$, respectively, we get nice collapse as shown in Fig.~\ref{fig:kz}(c2). Thus, this dynamical scaling collapse provides independent confirmation of the critical exponents extracted from the static analysis.

\textcolor{red}{\section{\label{sec:fisher}Application In Quantum Sensing}}

Quantum sensing aims to estimate
physical parameters with exceptional accuracy, and progress
has been made by exploiting diverse many-body phenomena, such as criticality near phase transition. The probe state and the measurement strategy play central roles in determining how effectively a physical parameter can
be estimated. In quantum-enhanced parameter estimation, the optimal measurement strategy utilizes quantum resources like a probe state being at or near a
quantum phase transition.  Carefully designed strategies can surpass the classical precision bound, known as the standard quantum limit (SQL), and approach, or even attain, the ultimate bound set by the quantum Cram\'er–Rao (CR) inequality \cite{cramer1999mathematical,braunstein1994statistical} that sets the precision limit. From the quantum CR bound, the uncertainty in estimating $\alpha$ over $M$ number of measurements satisfies $\delta\alpha^2\geq (M F_Q)^{-1}$ , where $F_Q$ is the  Quantum Fisher Information (QFI). It is evident that higher QFI is instrumental to achieve negligible uncertainty. In  the following, we will assess the performance of the parameter estimator using adiabatic and dynamic protocols. 

\subsection{\textnormal{\textbf{ Adiabatic Scaling}}}

We adiabatically shift $\alpha$ which is encoded in $\psi(\alpha)$. As we reach $\alpha_c$, transition can be captured  
by the fidelity susceptibility $\eta_Q$, which is defined as \cite{debnath2025tilt}
\begin{equation}
    \eta_Q(\alpha,\delta \alpha )=-\lim_{\delta \alpha \to 0} \frac{\partial^2 \mathcal{F}(\alpha, \delta \alpha)}{\partial (\delta \alpha)^2},    
\end{equation}
where, $\mathcal{F}(\alpha, \delta \alpha)=\langle\psi(\alpha)|\psi(\alpha+\delta\alpha)\rangle$ quantifies the overlapping amplitude between the ground state wave functions at $\alpha$ and at $\alpha+\delta \alpha$, where $\delta \alpha$ corresponds to a small shift. $\alpha_c$ represents the boundary across which there is a drastic change in the wavefunction structure in finite systems: it is localized in the regime $|\alpha| > |\alpha_c|$ and delocalized in the regime $|\alpha| < |\alpha_c|$. As a result the wavefunctions across the phases has small overlap. In particular, the fidelity susceptibility is directly proportional to the QFI, $F_Q(\alpha,\delta \alpha ) = 4\eta_Q(\alpha,\delta\alpha)$, a quantity that characterizes the ultimate precision limits allowed by quantum mechanics when estimating an unknown parameter. Recently, critical enhancement of the QFI has been widely recognized as a resource in quantum metrology and sensing, where it enables improved parameter-estimation precision  \cite{bayat20251,agarwal2025critical}. $F_Q$ can reflect the rapid variation of physical properties in the vicinity of a quantum critical point. Because of this, $F_Q$ is useful not only for identifying quantum phase transitions but also for developing quantum sensors that operate near criticality.

\begin{figure}
\centering
\includegraphics[scale=0.6]{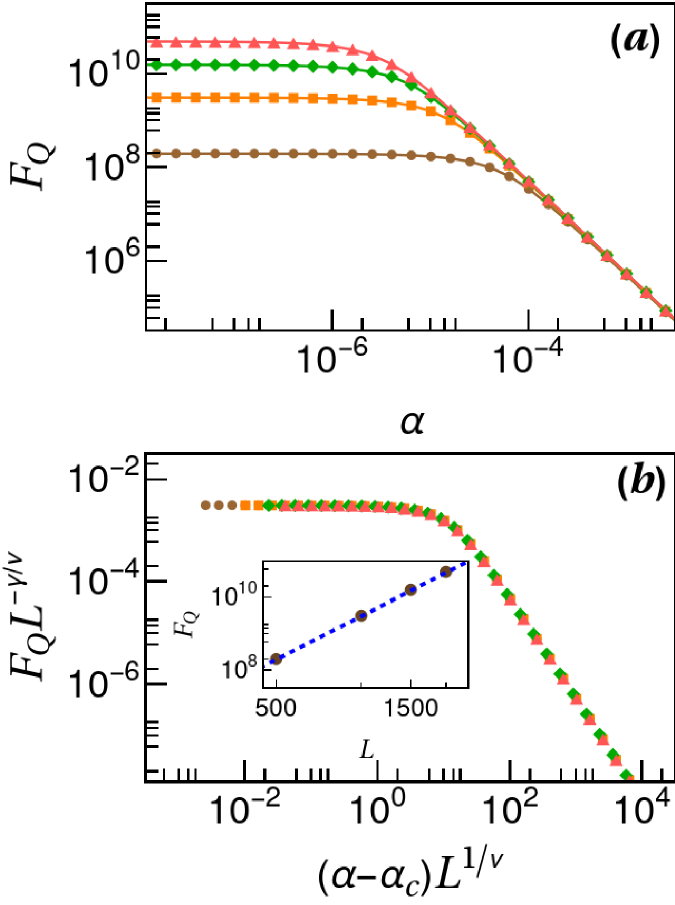}
\caption{(a) presents the QFI, $F_Q$ versus  $\alpha$ for different system sizes, $L=500$ (yellow circle), 1000 (orange square), 1500 (green diamond), 2000 (pink up-triangle) for ground state using exact diagonalization. (b) shows a collapse plot for the $F_Q$ with $\{\gamma, \nu\}\sim\{2.02(2), 0.49(1)\}$. In the inset we plot the
QFI, $F_Q$ (brown dots), as a function of
$L$ for the ground at $\alpha=10^{-8}$ and we fitted it with function $\propto L^\beta$ (blue dashed line). We found $\beta=3.98$.}\label{fig:fisher_all}
\end{figure}

In Fig.~\ref{fig:fisher_all}(a) we look into variation of QFI, $F_Q$, as a function of $\alpha$ for various $L$. As one may expect by now, the $F_Q$ remains nearly flat up to the finite threshold value, $\alpha_T$. The behavior is quite analogous with $\xi$. It is important to note that maximum value of $F_Q$ increases with $L$ as one lowers $\alpha$. We propose the following ansatz for extracting the associated scaling exponents \cite{debnath2025tilt},
\begin{eqnarray}
F_Q=L^{\gamma/\nu}f_7\left((\alpha-\alpha_c) L^{1/\nu}\right),\label{eq:six}
\end{eqnarray}
where $f_7[.]$ is an arbitrary function. Fig.~\ref{fig:fisher_all}(b) represents the best collapse plot that is in accordance to the cost function theory. The obtained scaling exponents are $\{\gamma,\nu\} = \{1.96(4),\,0.49(1)\}$

Apart from serving as a probe of quantum criticality, fidelity susceptibility also establishes a natural bridge to the concepts in quantum parameter estimation.  The sensing performance is determined by the finite size scaling of the QFI with $L$ \cite{debnath2025tilt}, $F_Q \sim L^{\beta}$. $\beta = 1$ is  known as SQL, which is the best that $L$ independent qubits can achieve \cite{giovannetti2004quantum,giovannetti2006quantum}, and a quantum-enhanced sensing implies $\beta > 1$. Quantum criticality-based adiabatic quantum sensors that can surpass, $\beta=2$, the so-called Heisenberg limit,  have been previously reported in different scenarios \cite{mondal2024multicritical, sahoo2024enhanced,sahoo2025stark,debnath2025tilt,sahoo2025power}. Our analysis implies that $\beta=\gamma/\nu$, and $F_Q \sim L^{4}$. Therefore, the scaling behavior extracted from the QFI in our model not only reveals its critical properties but also signals potential applications in criticality-assisted quantum sensing platforms.

\begin{figure}
\centering
\includegraphics[scale=0.6]{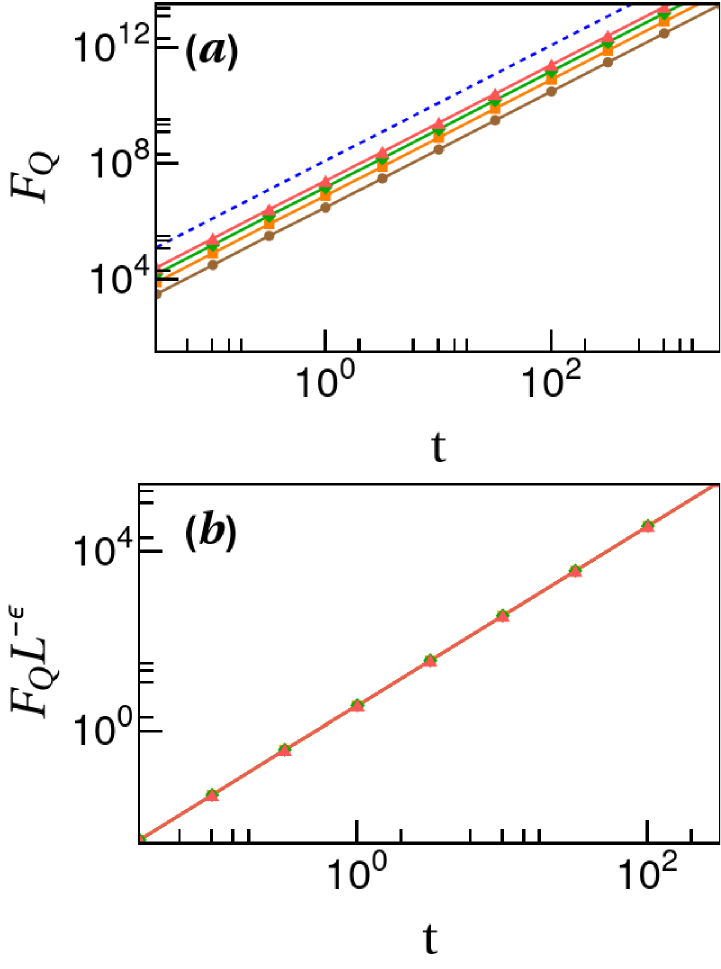}
\caption{(a) Variation of $F_Q$ with $t$ for different system sizes, $L=400$ (yellow circle), 600 (orange square), 800 (green diamond), 1000 (pink up-triangle). We take delocalized ground state at $\alpha=10^{-8}$ as initial state. The fit gives $F_Q$ scales with $t$ as $F_Q\sim t^2$ shown in blue dashed line. Figure (b) shows collapse plot after rescaling $F_Q$ with $L^\epsilon$ using $\epsilon=2.28$. This gives $F_Q\sim L^{2.28}t^{2}$.}\label{fig:dy_fisher}
\end{figure}

\subsubsection{Effects of state preparation time}

We critically evaluate the proposed sensing protocols to identify their limitations under realistic experimental constraints. In particular, perfect adiabaticity is an idealization; in practice, parameter variations necessarily occur over finite timescales. If the time $t$ is required for ground state to reach the target state adiabatically, we can connect it to energy gap as $t\sim \hbar/{\Delta E}$ \cite{sahoo2024enhanced}. Recalling the relation at near criticality
$\Delta E\sim L^{-z}$, we can write $t\sim L^z$. We can conclude that probe state preparation time depends on system size and scaling exponent related to energy gap. This modifies QFI to $F_Q/t\sim L^{\beta-z}$. Recalling our $z$ and $\beta$ values, we get $F_Q/t\sim L^{2}$. Even though inclusion of state preparation time reduces QFI scaling, it still promises quantum advantage in sensing.

\subsection{\textnormal{\textbf{Dynamic Scaling}}}

We now consider a sensing protocol based on nonequilibrium dynamics. Its operating principle relies on the enhanced susceptibility of a quantum system in the vicinity of a quantum phase transition, where the ground state undergoes pronounced changes even under small variations of the Hamiltonian parameters. In this protocol, the system is initially prepared in its ground state and after using sudden quench we move the system out of equilibrium across the critical point. The resulting time evolution carries information about the unknown parameter through the system's dynamical susceptibility.  Here the protocol duration time $t$ manifests itself as a fundamental resource
for the QFI scaling, along with the system size $L$. Similar
to the adiabatic considerations, sensing protocols have been
designed via non-equilibrium dynamics  influenced
by criticality corresponding to a second-order quantum phase
transition \cite{tsang2013quantum,mishra2021driving,chu2021dynamic,garbe2022critical}. In particular, under certain assumptions, the dynamical precision quantified via QFI can at
most scale as $L^{2}t^{2}$ , where $t$ represents total interrogation time \cite{boixo2007generalized,garbe2020critical,ilias2022criticality}. This is the so-called Heisenberg limit in the dynamical context \cite{giovannetti2004quantum,giovannetti2006quantum}.
The Fisher information can be defined like before as

\begin{equation}
    F_Q(\alpha,\delta \alpha,t )=-4\lim_{\delta \alpha \to 0} \frac{\partial^2 \mathcal{F}(\alpha, \delta \alpha, t)}{\partial (\delta \alpha)^2},    
\end{equation}
where, $\mathcal{F}(\alpha, \delta \alpha, t)=\langle\psi(\alpha,t)|\psi(\alpha+\delta\alpha,t)\rangle=\langle\psi_0|e^{-i H(\alpha)t}e^{-i H(\alpha+\delta\alpha)t}|\psi_0\rangle$ is the overlap
between two dynamical states separated by a perturbation $\delta\alpha$. An extended phase at $\alpha_i=10^{-8}$ is quenched in the localized phase at $\alpha_f=1$
through the transition point: $e^{-i H(\alpha_f)t}|\psi(\alpha_i)\rangle$. We show the dependence of QFI on $t$ for several system sizes in Fig.~\ref{fig:dy_fisher}(a). $F_Q$ scales with $t$ as $F_Q\sim t^2$ (blue dashed line). Data collapse suggests $F_Q$ scales as $L^{2.28}t^{2}$ shown in Fig.~\ref{fig:dy_fisher}(b). The ultimate precision of our system can be determined from the semi-norm of the Hamiltonian $H$~\cite{boixo2007generalized}. In our case, we end-up with a scaling of QFI as $F_Q \sim L^{2(1+\alpha)}$.

\section{\texorpdfstring{\textcolor{red}{Discussion}}{Discussion}\label{sec:conc}}

To summarize, in this work, we investigate localization phenomena in a one-dimensional lattice with a power-law graded hopping amplitude. This model realizes a novel kinetic route to localization, distinct from the disorder, quasi-periodic, or potential-driven mechanisms characterizing Anderson, Aubry–Andr{\'e}, and Stark localization. In the thermodynamic limit the system becomes localized at $|\alpha| \to 0$, signaling the presence of a critical point characterized by a diverging localization length. Through a finite-size scaling analysis of localization length, inverse participation ratio, fidelity susceptibility, and spectral-gap, we report the associated scaling exponents revealing a new universality class. In the one-dimensional Anderson model with weak diagonal disorder, the band-edge localization length is known to diverge with the exponent $\nu=2/3$ \cite{cestari2011critical}, whereas in the present graded-hopping model our finite-size scaling gives $\nu=1/2$. Thus the distinction is not only microscopic but also reflected in the scaling properties.

Finally, we performed complementary non-equilibrium ramp dynamics in order to further exhibit Kibble–Zurek scaling, that showed that the dynamical critical behavior can be consistently described with the static exponents. This work proposes to use localization-inducing  power-law graded hopping that can enhance the precision of parameter estimation. Near the localization transition, a small change in $\alpha$
produces a large deformation of the spatial wavefunction, leading to an enhanced QFI. This enhancement
would not arise from an incoherent classical probability distribution in the same way; it relies on the
quantum coherence of the single-particle wavefunction across lattice modes. We emphasize that the origin of enhanced QFI does not stem from multipartite entanglement or many-body interactions. This differs from conventional entanglement-based quantum sensing schemes, where quantum enhancement is achieved through multipartite entangled states such as GHZ or other many-body correlated states \cite{frowis2011stable}. However, those schemes are
prone to decoherence and are fundamentally limited to
Heisenberg scaling.

On the applied side, the sensitivity of the system near criticality suggests potential relevance for quantum metrology and sensing. We establish a coherent picture of static and dynamic scaling in our model, while also underlining the dual role of criticality as both a diagnostic tool and a practical resource. Our protocol should be viewed as a critical metrological protocol for engineered hopping profiles in programmable quantum-simulation platforms. The sensing task is to detect or calibrate small deviations in a spatially engineered kinetic landscape. This is experimentally meaningful in several platforms. In optical cold-atom systems, spatially structured laser fields and Raman couplings provide a route to programmable lattice Hamiltonian \cite{masi2021spatial,salamon2020simulating}. 
The most direct way to get $t_i=i^{\alpha}$ within optical lattice set-up is to use a Digital Micromirror Device (DMD) or a Spatial Light Modulator (SLM) to project a highly customized, point-by-point potential landscape \cite{Zupancic16,Gauthier16}. These illustrates the degree of spatial programmability available in cold-atom platforms. Apart from that, in photonics, spatially varying the evanescent coupling between neighboring waveguides another mature method for engineering custom, non-uniform tight-binding distributions \cite{Izdebskaya11, Szameit2010}. Trapped-ion systems could also provide another possible setting, since effective site- or bond-dependent couplings can also be engineered there \cite{smith2016many,rajabi2019dynamical,morong2021observation}.

Exploring analytic continuum descriptions and real-space field-theoretical descriptions may lead to a deeper understanding of the emergent effective potentials responsible for the observed scaling behavior. Similar analysis beyond single particle with interaction can be next interesting thing to look into.

\appendix

\section{Continuous limit}
To obtain the continuum envelope form, we denote the bond hopping by
 $t_{i+1/2}=t(x_i+a/2)$, where $x_i=ia$ (considering the fact that $t$ lives on the bond, where as $\psi$ is defined on the site, and $a$ is the lattice constant). The single-particle
 eigenvalue equation is
\begin{equation}
 E\psi_i=-t_{i+1/2}\psi_{i+1}-t_{i-1/2}\psi_{i-1}.
\end{equation}
We use the Taylor expansions
$t_{i\pm 1/2}
=
t(x)\pm \frac{a}{2}t'(x)+\mathcal{O}(a^2)$,
and
$\psi_{i\pm 1}
=
\psi(x)\pm a\psi'(x)+\frac{a^2}{2}\psi''(x)+\mathcal{O}(a^3).$
Few algebraic manipulation then leads us to the continuum envelope Hamiltonian as,
\begin{equation}
H_{\rm env}
=
-2t(x)
-a^2\frac{\partial}{\partial x}
\left[
t(x)\frac{\partial}{\partial x}
\right],
\end{equation}
up to higher-gradient corrections. For the graded hopping model,
$t(x)=J(x/a)^\alpha$, and in units $J=a=1$, $t(x)=x^\alpha$, from which the final resulting form of the asymptotic equation in continuum turns out to be
\begin{equation}
    \label{asymptotic}
    \psi''(x)+\frac{\alpha}{x}\psi'(x)+(2+Ex^{-\alpha}) \psi(x)=0.
\end{equation}
This equation have analytical solution for specific choices $\alpha$, such as $\alpha=0, 1, 2$. However, in general, it does not admit a closed form solution, as per our knowledge. The numerically obtained ground state eigenfunction of Eq.~\ref{asymptotic} agrees with the results from the direct tight-binding diagonalization from Eq.~\ref{eq:main}. 

However, the continuous limit looses crucial information on lattice periodicity. This distinction is physically important. The continuum extension does not retain the lattice structure of the tight-binding model, and information originating from periodicity, such as finite bandwidth, Brillouin-zone periodicity, or lattice-induced interference effects. Thus, although continuum analysis is useful for asymptotic insight, but it may not be most suitable for full tight-binding treatment of many involved physics of the tight-binding systems, particularly in context of the dynamical studies.

\section*{Data availability statement}

The data that support the findings of this study are available from the corresponding author upon reasonable request.

\section*{Conflict of interest}

The authors declare no financial or organizational conflicts of interest related to this work.

\bibliographystyle{unsrt}
\bibliography{apssamp}

\end{document}